\documentclass[twocolumn]{aastex701}

\newcommand{\tefft}{$T_{\mbox{\scriptsize eff}}$} 
\newcommand{\logg}{\ensuremath{\log g}}

\usepackage{subfigure}
\usepackage{ textcomp }
\usepackage{ tipa }
\usepackage{ amssymb }
\usepackage{amsmath}
\usepackage{ wasysym }
\usepackage{CJK}
\usepackage{threeparttable}
\usepackage{booktabs}
\usepackage{newtxtext,newtxmath}
\usepackage[T1]{fontenc}
\usepackage{amsmath}	
\usepackage{graphicx}
\def\kms{{\rm\,km\,s^{-1}}}

\def\eg{{ e.g.,\ }}

\def\lta{\mathrel{\spose{\lower 3pt\hbox{$\mathchar"218$}}
     \raise 2.0pt\hbox{$\mathchar"13C$}}}
\def\gta{\mathrel{\spose{\lower 3pt\hbox{$\mathchar"218$}}
     \raise 2.0pt\hbox{$\mathchar"13E$}}}

\def\feh{[Fe/H]}

\makeatletter

\newcommand{\Rmnum}[1]{\expandafter\@slowromancap\romannumeral #1@}
\newcommand{\teff}{$T_{\rm eff}$}

\makeatother

\def\vt{$\xi_{\rm t}$}

\defcitealias{Roederer+2014AJ}{R14}
\defcitealias{Li+2018ApJS..238...16L}{L18}
\defcitealias{Ceccarelli+2025arXiv251006332C}{C25}
\defcitealias{Yuan+2020}{Y20}

\begin{document}

\title{HR-GO II: chemical abundances of low-$E$ retrograde dynamically-tagged-groups: \\ Revealing Thamnos as a very metal-poor substructure}

\correspondingauthor{Haining Li}
\email{lhn@nao.cas.cn}
\correspondingauthor{Zhen Yuan}
\email{zhen.yuan@nju.edu.cn}

\author[0009-0002-4282-668X]{Renjing Xie}
\affiliation{CAS Key Laboratory of Optical Astronomy, National Astronomical Observatories, Chinese Academy of Sciences, Beijing 100101, China}
\affiliation{School of Astronomy and Space Science, University of Chinese Academy of Sciences, No.19(A) Yuquan Road, \\ Shijingshan District, Beijing, 100049, China}
\email{xierj@bao.ac.cn}

\author[0000-0002-8129-5415]{Zhen Yuan}
\affiliation{School of Astronomy and Space Science, Nanjing University, Nanjing 210093, China}
\affiliation{Key Laboratory of Modern Astronomy and Astrophysics (Nanjing University), Ministry of Education, Nanjing 210093, China}
\email{zhen.yuan@nju.edu.cn}

\author[0000-0002-0389-9264]{Haining Li}
\affiliation{CAS Key Laboratory of Optical Astronomy, National Astronomical Observatories, Chinese Academy of Sciences, Beijing 100101, China}
\email{lhn@nao.cas.cn}

\author[0000-0002-8077-4617]{Tadafumi Matsuno}
\affiliation{Astronomisches Rechen-Institut, Zentrum f\"ur Astronomie der Universit\"at Heidelberg, M\"onchhofstra{\ss}e 12-14, 69120 Heidelberg, Germany}
\email{matsuno@uni-heidelberg.de}

\author[0000-0002-1349-202X]{Nicolas F. Martin}
\affiliation{Universit\'e de Strasbourg, CNRS, Observatoire astronomique de Strasbourg, UMR 7550, F-67000 Strasbourg, France\\
Max-Planck-Institut f\"{u}r Astronomie, K\"{o}nigstuhl 17, D-69117 Heidelberg, Germany \\}
\email{nicolas.martin@astro.unistra.fr}

\author[0009-0008-1319-1084]{Ruizhi Zhang}
\affiliation{CAS Key Laboratory of Optical Astronomy, National Astronomical Observatories, Chinese Academy of Sciences, Beijing 100101, China}
\affiliation{Instituto de Astrofísica de Canarias, Vía Láctea S/N, E-38205 La Laguna, Tenerife, Spain}
\affiliation{Universidad de La Laguna, Departamento de Astrofísica, E-38206 La Laguna, Tenerife, Spain}
\email{ruizhi.zhang-ext@iac.es}

\author[0000-0001-7395-1198]{Zhiqiang Yan}
\affiliation{School of Astronomy and Space Science, Nanjing University, Nanjing 210093, China}
\affiliation{Key Laboratory of Modern Astronomy and Astrophysics (Nanjing University), Ministry of Education, Nanjing 210093, China}
\email{yan@nju.edu.cn}

\author[0000-0002-3182-3574]{Federico Sestito}
\affiliation{Centre for Astrophysics Research, Department of Physics, Astronomy and Mathematics, University of Hertfordshire, Hatfield AL10 9AB, UK}
\email{f.sestito@herts.ac.uk}

\author[0000-0002-2468-5521]{Guillaume F. {Thomas}}
\affiliation{Instituto de Astrofísica de Canarias, Vía Láctea S/N, E-38205 La Laguna, Tenerife, Spain}
\affiliation{Universidad de La Laguna, Departamento de Astrofísica, E-38206 La Laguna, Tenerife, Spain}
\email{guillaume.thomas.astro@gmail.com}

\author[0000-0002-6389-2697]{Projjwal Banerjee}
\affiliation{Indian Institute of Technology Palakkad, Kerala, India}
\email{projjwal.banerjee@gmail.com}

\author[0009-0001-0604-072X]{Ruizheng Jiang}
\affiliation{CAS Key Laboratory of Optical Astronomy, National Astronomical Observatories, Chinese Academy of Sciences, Beijing 100101, China}
\affiliation{School of Astronomy and Space Science, University of Chinese Academy of Sciences, No.19(A) Yuquan Road, \\ Shijingshan District, Beijing, 100049, China}
\email{jiangrz@bao.ac.cn}

\author[0000-0001-5122-7422]{Linda Lombardo}
\affiliation{INAF, Osservatorio Astronomico di Trieste, Via Tiepolo 11, 34143 Trieste, Italy}
\email{linda.lombardo@inaf.it}

\author[0000-0001-5200-3973]{David S. Aguado}
\affiliation{Instituto de Astrofísica de Canarias, Vía Láctea S/N, E-38205 La Laguna, Tenerife, Spain}
\affiliation{Universidad de La Laguna, Departamento de Astrofísica, E-38206 La Laguna, Tenerife, Spain}
\email{david.aguado@iac.es}

\author[0000-0001-6924-8862]{Kohei Hattori}
\affiliation{National Astronomical Observatory of Japan, 2-21-1 Osawa, Mitaka, Tokyo 181-8588, Japan}
\affiliation{The Institute of Statistical Mathematics, 10-3 Midoricho, Tachikawa, Tokyo 190-8562, Japan }
\email{kohei.hattori@nao.ac.jp}

\author[0000-0002-8980-945X]{Gang Zhao}
\affiliation{CAS Key Laboratory of Optical Astronomy, National Astronomical Observatories, Chinese Academy of Sciences, Beijing 100101, China}
\affiliation{School of Astronomy and Space Science, University of Chinese Academy of Sciences, No.19(A) Yuquan Road, \\ Shijingshan District, Beijing, 100049, China}
\email{gzhao@nao.cas.cn}

\begin{abstract}

Milky Way halo substructures identified in dynamical space are known to suffer from contamination from the Milky Way in-situ stars, which makes their accreted origins uncertain. We present detailed chemical abundances of 35 stars belonging to two sets of dynamically tagged groups, Rg8 and Rg9, to investigate their accreted nature. Both groups are composed of stars with low orbital energy and very retrograde orbits. We find that Rg8 and Rg9 are chemically indistinguishable across all elements, from C to Eu, strongly indicating that they belong to the same structure. The iron-abundance distribution of this low-$E$ retrograde group has a prominent peak at [Fe/H] $\approx-2.1$, revealing that its main population is very metal-poor, and a secondary peak at [Fe/H] $\approx-1.5$, very likely due to contamination from Milky Way in-situ stars. 
These groups also heavily overlap with the Thamnos substructure in dynamical space, and we thus use them to investigate the chemical properties of Thamnos. The dominant, low-metallicity population provides strong evidence for the ex-situ origin of Thamnos, as well as its very metal-poor nature. We do not see any evidence of an $\alpha$ knee in our sample, which is consistent with previous studies.
Comparison with the Cetus-Palca stream in the chemical space shows similar abundance distributions, and thus it suggests that the Thamnos progenitor dwarf galaxy had a truncated star formation history due to its early merger with the Milky Way. 


\end{abstract}


\section{Introduction}
\label{sec:intro}
Stars lacking metals have long been regarded as fossil records of the early evolution of the Galaxy and preserve important clues about the first generation of stars in the universe \citep{Freeman+2002ARA&A..40..487F,Beers+2005,Frebel+2015}. 
The detailed chemical abundance distributions of these objects can be compared to the yields of the first generation of stars to constrain their nucleosynthesis \citep{Nomoto+2013,Kobayashi+2020}, while their observed abundance trends with metallicity provide essential information about the chemical history of galaxies \citep{Tolstoy+2009}.

According to the $\Lambda$CDM cosmology \citep{Springel+2005Natur.435..629S}, galaxies assemble hierarchically with time, growing their halos through mergers and the accretion of smaller stellar systems \citep{Searle+1978ApJ...225..357S,White+1991ApJ...379...52W,Bullock+2005ApJ...635..931B}. The sequence of these events leaves its imprint on the distribution of stars in the host galaxy \citep{Helmi+2008A&ARv..15..145H,Helmi+2020}. The identification and the characterisation of debris within our own Galaxy's halo offers insights into the significance of accretion and the process of galaxy formation and evolution.

Leveraging Gaia data \citep{Gaia+2023,Katz+2023} alongside photometric and spectroscopic information from large sky surveys --- including the Sloan Digital Sky Survey 
\citep[SDSS;][]{Abazajian+2009ApJS..182..543A}, the Apache Point Observatory Galactic Evolution Experiment 
\citep[APOGEE;][]{Majewski+2017}, the Large Sky Area Multi-Object Fibre Spectroscopic Telescope
\citep[LAMOST;][]{Zhao+2006,zhao+2012}, and the Galactic Archaeology with HERMES survey 
\citep[GALAH;][]{Silva+2015} --- has revealed a wealth of dynamical substructures within the Galactic stellar halo, including the Gaia-Sausage/Enceladus 
\citep[GSE;][]{Belokurov+2018,Haywood+2018,Helmi+2018}, 
Sequoia
\citep{Myeong+2019,Matsuno+2019},
Thamnos 
\citep{Koppelman+2019},
Arjuna
\citep{Naidu+2020ApJ...901...48N},
I'itoi
\citep{Naidu+2020ApJ...901...48N},
LMS-1/Wukong \citep{Yuan+2020ApJ...898L..37Y,Naidu+2020ApJ...901...48N}, Pontus \citep{Malhan2022ApJL},
Shiva \citep{Malhan2024ApJ},
and Shakti \citep{Malhan2024ApJ}
structures, as well as the inner Galactic substructure Koala/Kraken/Heracles
\citep{Forbes+2020MNRAS.493..847F,Kruijssen+2020MNRAS.498.2472K,Horta+2021}.

In our previous work, we call substructures identified in orbital space dynamically-tagged groups (DTGs) instead of substructures \citep{Yuan+2020}.
This is because clumps and clusters identified from dynamical space could come from the same accreted system \citep{Amarante+2022ApJ...937...12A, Khoperskov2023, Thomas+2025arXiv250410398T} or, as shown by several numerical simulations, they could be created by stars formed in the Milky Way disk (in-situ stars) that were propelled on halo-like orbits as a response to past mergers 
\citep{Gmez+2012MNRAS.423.3727G,Jean+2017A&A...604A.106J,Laporte+2018MNRAS.481..286L,Thomas2019ApJ...886...10T,Thomas+2025arXiv250410398T}.

To investigate the origin of these DTGs, chemical abundances of several of them have been studied via high-resolution (HR) spectroscopy, including GSE \citep{Aguado+2021,Carrillo+2022,Xie+2025ApJ...985..250X}, the Helmi stream \citep{Matsuno+2022A&A...665A..46M,Gull+2021ApJ...912...52G,Aguado+2021MNRAS.500..889A}, Sequoia \citep{Aguado+2021, Matsuno+2022Seq}, Wukong/LMS-1 \citep{Limberg+2024}, and Thamnos \citep{Ceccarelli+2025arXiv251006332C}. Comparative studies covering multiple substructures \citep{Horta+2023,Zhang+2024} and the retrograde halo \citep{Ceccarelli+2024A&A...684A..37C} have also been performed.

Although previous works suggested that Thamnos exhibits distinct chemical properties compared to other substructures \citep[e.g.,][]{Koppelman+2019, Horta+2023}, a comprehensive understanding has been hampered by sample limitations --- either focusing solely on the very metal-poor regime \citep[e.g.,][]{Zhang+2024} or including a limited range of chemical species \citep{Monty+2020}.
Additionally, a recent study on the elemental abundances of hundreds of Thamnos candidates reported significant contamination from in-situ Milky Way stars \citep[over 70 \%;][]{Ceccarelli+2025arXiv251006332C}, making it more complicated to reveal the true nature of Thamnos.
Whether it is a bona-fide remnant from an accreted dwarf galaxy or simply a set of overdensities arising from in-situ MW stars remains unclear.


In this work, we use HR spectroscopy to explore the chemical-abundance signature of two sets of groups, Rg8 and Rg9, originally identified by \citet{Yuan+2020} as DTG-21,24,29 and DTG-22,28-33, respectively, from the LAMOST catalogue of very metal-poor stars (VMP; \citealt{Li+2018ApJS..238...16L}). These two groups have very retrograde motion and low orbital energy, similar to the Thamnos stars. However, compared to the more metal-rich regime, the low metallicity regime in which these groups were identified is expected to be less contaminated by in-situ Milky Way stars. Therefore, clumps and clusters identified from VMP samples are more likely to be genuine accreted structures.

To investigate these two sets of DTGs without metallicity constraints, we further expand the member lists by using the \texttt{StarGO} algorithm \citep{Yuan+2018ApJ...863...26Y} on the \emph{Gaia} RVS sample \citep{Gaia+2016A&A...595A...1G,Gaia+2018A&A...616A...1G}. In total, we followed up 35 stars with HR spectrographs using telescopes from both hemispheres to decipher the chemical origins of these low-$E$ retrograde DTGs. The selection of targets and the observations are described in detail in Sec.~\ref{sec:obs}. The methods to derive stellar parameters and the abundance determination are explained in Sec. \ref{para}. We show the resulting detailed chemical abundances in Sec. \ref{sec:abund} and we discuss the associations between these DTGs and Thamnos in Sec. \ref{sec:result}, before summarizing our conclusions in Sec. \ref{sec:con}.

This is the second paper in the High-Resolution spectroscopic program on the Galactic Origins of elements (HR-GO). The HR-GO program focuses on stars in stellar streams and substructures that are stripped from accreted dwarf galaxies. The goal is to understand the production channels of elements in dwarf galaxies as well as their evolution histories by observing stars in their debris. The first work of the HR-GO series is the comprehensive elemental abundance study of the Cetus-Palca stream \citep{Sitnova+2024A&A...690A.331S}, which was produced by a disrupted low-mass dwarf galaxy. The Cetus sample also offers an important sample to compare our Rg8 and Rg9 results to.

\section{Target Selection and Observations}

\begin{figure*}
\plotone{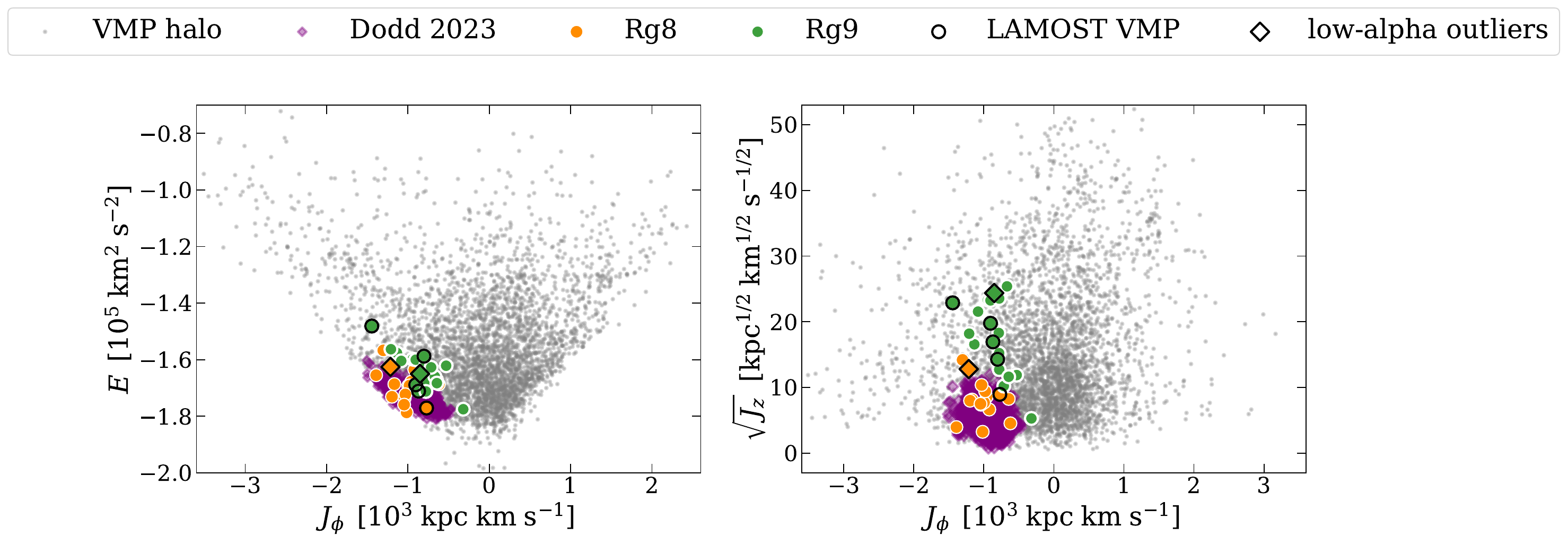}
\caption{Distributions of the Rg8 and Rg9 stars in energy and action space, along with stars used for comparison. The gray dots represent the halo sample from the LAMOST DR3 VMP catalogue \citep{Li+2018ApJS..238...16L}, updated with Gaia DR3 kinematics. Purple diamonds are Thamnos members from \citet{Dodd+2023A&A...670L...2D}, shown as a reference sample. The two groups studied in this paper are represented by orange filled circles (Rg8) and green filled circles (Rg9). Additionally, black open circles denote specific stars selected from the LAMOST VMP catalog, and the black open diamonds mark low-Mg outliers.}
\label{fig:dynamics}
\end{figure*}

\label{sec:obs}
We first select stars from the member lists of Rg8 (DTG-21,24,29) and Rg9 (DTG-22,28-33), identified as two very retrograde groups by \citet{Yuan+2020}, using the LAMOST DR3 VMP catalogue \citep{Li+2018ApJS..238...16L}. This name convention follows that of retrograde groups proposed by \citet{myeong2018}.  We consider Rg8 and Rg9 as two distinct dynamical groups because Rg8 has a slightly lower orbital energy and smaller vertical motion ($J_{\mathrm z}$) compared to Rg9, as shown in Fig.~\ref{fig:dynamics}. Whether they have different origins or not requires a detailed chemical abundance analysis. We also compare Rg8 and Rg9 with a prominent retrograde substructure, 
the Thamnos, using the selection of \citet{Dodd+2023A&A...670L...2D} in the dynamical space shown in Fig.~\ref{fig:dynamics}. Thamnos stars have relatively low $E$ and $J_{\mathrm z}$ and Rg8 stars mostly overlap them. The relationship between these two groups will be discussed in Sec. \ref{sec:result}.

From the DTG catalogue of VMP stars identified by \citet{Yuan+2020}, there are 39 stars brighter $G = 13.5$ out of a total of 208 for Rg8 and Rg9. In order to search for potential candidate members of these retrograde DTGs at higher metallicity, we utilize the \texttt{StarGO} \citep{Yuan+2018ApJ...863...26Y} to identify stars with similar dynamical properties in the $Gaia$ DR2 RVS sample \citep{Gaia+2018A&A...616A...1G,Katz+2019A&A...622A.205K}. This approach allows us to identify new members that are dynamically associated with these retrograde DTGs and also bright enough for HR followups. The membership identification has been proved to be robust when using the same approach to identify new members in the Cetus \citep{Yuan+2022ApJ...930..103Y} and C-19 \citep{Yuan+2025A&A...698A..82Y} streams. 

Note that all follow-up observations were performed before the release of \emph{Gaia} DR3 \citep{Gaia+2023}, when there was no metallicity information for the RVS stars. In this work, we update their kinematics using the \emph{Gaia} DR3 data and confirm that their membership remains valid. In total, we followed up 35 bright stars ($G<$ 13.5) associated with Rg8 and Rg9, including 6 VMP stars originally identified from the LAMOST DR3 VMP catalogue and 29 stars from the \emph{Gaia} DR2 RVS sample.

For each target, Table~\ref{tab:obser} provides information on the telescope and HR spectrograph used, the exposure time, and the resulting signal-to-noise ratio (S/N) in the wavelength region around $\lambda \sim 6000$ \AA. All spectra in the sample studied here were observed with four telescopes from both hemispheres:

\begin{itemize}
    \item the High Dispersion Spectrograph \citep[HDS,][]{Noguchi+2002PASJ...54..855N} on Subaru in 2020B (PI: Matsuno);
    \item the Magellan Inamori Kyocera Echelle spectrograph \citep[MIKE,][]{Bernstein+2003SPIE.4841.1694B} in 2019B (PI: Hatrori);
    \item the Echelle SpectroPolarimetric Device for the Observation of Stars \citep[ESPaDOnS][]{Donati+2006} on the Canada–France–Hawaii Telescope (CFHT) in 2021A (PI: Yuan);
    \item the UV-Visual Echelle Spectrograph \citep[UVES,][]{Dekker+2000SPIE.4008..534D} (Program ID: 106.21M6.001) on the Very Large Telescope (VLT) in 2021B (PI: Martin).
\end{itemize}

The HDS spectra are observed using the standard StdYd setup, which provides a wavelength coverage of 4000--5340 and 5450--6800 \AA, a resolving power R = $\Delta \lambda/\lambda = 45\,000$. The data is reduced using the IRAF\footnote{IRAF is distributed by the National Optical Astronomy Observatory, which is operated by the Association of Universities for Research in Astronomy (AURA) under a cooperative agreement with the National Science Foundation} script hdsql\footnote{http://www.subarutelescope.org/Observing/Instruments/HDS/hdsql-e.html} that includes CCD linearity correction, scattered light subtraction, aperture extraction, flat-fielding, wavelength calibration, and heliocentric velocity correction. 

The MIKE spectra are taken with the $0.7''$ slit width, which yields  R $\sim 28\,000$ and $\sim 35\,000$ in the red and blue wavelength regions, respectively. The combined wavelength coverage is 3300--9600 \AA. Data reduction is carried out with the MIKE Carnegie Python pipeline \citep{Kelson+2003PASP..115..688K}. 

The UVES spectra are observed with the $1.0''$ slit width, which provides R $\sim 40\,000$. The wavelength coverage of the three arms is 3750--5000, 5700--7500, and 7660--9450 \AA. The observed spectra are reduced with the ESO pipeline\footnote{https://www.eso.org/sci/software/pipelines/}.

The ESPaDOnS spectra were obtained as part of program 20BF006. The observations provide a wavelength coverage from 3600 to 10\,800\,\AA, with a resolving power of $R \sim 68\,000$. The data reduction was performed with the dedicated \textsc{esprit} pipeline\footnote{\url{https://www.cfht.hawaii.edu/Instruments/Spectroscopy/Espadons/Espadons_esprit.html}}.

\begin{deluxetable*}{ccccccllcccccccccc}
\tabletypesize{\scriptsize}
\label{tab:obser}
\tablewidth{0pt} 
\tablecaption{Stellar sample, stellar parameters, and characteristics of the observed spectra. \label{tab:obser}}
\tablehead{
\colhead{short ID} &\colhead{Gaia ID} & \colhead{RA} & \colhead{Dec} & \colhead{G} & \colhead{$t_{\rm exp}$} & \colhead{Tel./Inst.} & \colhead{S/N} & \colhead{\teff} & \colhead{\teff(err)} & \colhead{\logg} & \colhead{\logg(err)} & \colhead{\feh} & \colhead{\feh(err)} & \colhead{\vt} & \colhead{\vt(err)} & \colhead{source} & \colhead{group}\\
\colhead{} & \colhead{} & \colhead{(deg)} & \colhead{(deg)} & \colhead{} 
& \colhead{(s)} & \colhead{} & \colhead{per pixel} & \colhead{(K)} & \colhead{(K)} & \colhead{(dex)} & \colhead{(dex)} & \colhead{(dex)} & \colhead{(dex)} & \colhead{($\kms$)} & \colhead{($\kms$)} & \colhead{}
} 
\startdata
         Rg9\_1 & 4972678230112467968 & 5.408557143396869 & -51.96999443073288 & 12.81 & 1350.0 & Magellan/MIKE & 133.4 & 5077.817 & 112.722 & 2.098 & 0.036 & -2.05 & 0.09 & 1.71 & 0.14 & RVS & Rg9 \\ \hline
        Rg9\_2 & 4711709553278850048 & 25.9993656873449 & -63.24083067788601 & 11.21 & 750.0 & Magellan/MIKE & 165.4 & 4833.173 & 94.892 & 1.881 & 0.028 & -1.63 & 0.1 & 1.74 & 0.13 & RVS & Rg9 \\ \hline
        Rg8\_1 & 4851489027705355776 & 49.118785780916056 & -40.60319633287946 & 11.97 & 3600.0 & VLT/UVES & 94.2 & 4974.713 & 102.089 & 2.205 & 0.03 & -2.0 & 0.08 & 1.65 & 0.09 & RVS & Rg8 \\ \hline
        Rg9\_3 & 2902434449329878144 & 84.30434352180205 & -31.22200639982802 & 11.77 & 2880.0 & VLT/UVES & 65.8 & 4972.573 & 105.646 & 2.014 & 0.03 & -2.17 & 0.11 & 1.75 & 0.14 & RVS & Rg9 \\ \hline
        Rg8\_2 & 5267028900300303488 & 107.63009293049194 & -71.36415837061494 & 10.60 & 2160.0 & VLT/UVES & 87.9 & 5120.051 & 96.549 & 2.67 & 0.023 & -1.55 & 0.12 & 1.38 & 0.1 & RVS & Rg8 \\ \hline
        Rg9\_4 & 657021355093742848 & 123.77704298002514 & 17.605019757464262 & 11.03 & 2679.0 & CFHT/ESPaDOnS & 125.7 & 4627.606 & 91.972 & 1.674 & 0.036 & -0.9 & 0.09 & 1.83 & 0.12 & RVS & Rg9 \\ \hline
        Rg9\_5 & 651093372510623744 & 126.12361121852184 & 13.16992624765404 & 12.94 & 1900.0 & VLT/UVES & 105.45 & 5414.091 & 123.587 & 3.432 & 0.026 & -2.08 & 0.12 & 1.2 & 0.1 & LAMOST VMP & Rg9 \\ \hline
        Rg8\_3 & 5708148155001776512 & 126.15904372789976 & -17.923322200420596 & 9.95 & 2679.0 & CFHT/ESPaDOnS & 132.7 & 4565.67 & 83.186 & 1.176 & 0.029 & -2.22 & 0.1 & 2.0 & 0.12 & RVS & Rg8 \\ \hline
        Rg8\_4 & 582973163970364416 & 131.88380705256918 & 5.803893539483561 & 11.34 & 2679.0 & CFHT/ESPaDOnS & 141.0 & 5055.005 & 94.953 & 2.517 & 0.025 & -1.78 & 0.12 & 1.52 & 0.08 & RVS & Rg8 \\ \hline
        Rg9\_6 & 5219649806094112768 & 142.10092320488107 & -69.74121640632211 & 12.19 & 900.0 & Magellan/MIKE & 141.7 & 4963.32 & 90.405 & 2.049 & 0.028 & -2.24 & 0.09 & 1.77 & 0.17 & RVS & Rg9 \\ \hline
        Rg9\_7 & 3827038413057098624 & 144.48366079888515 & -2.7655723831280894 & 9.74 & 2679.0 & CFHT/ESPaDOnS & 129.4 & 5194.987 & 123.151 & 3.083 & 0.03 & -1.06 & 0.13 & 1.1 & 0.17 & RVS & Rg9 \\ \hline
        Rg9\_8 & 3769595523279885952 & 148.96253402276037 & -11.4093730018335 & 10.31 & 2679.0 & CFHT/ESPaDOnS & 129.7 & 4883.917 & 95.76 & 1.943 & 0.029 & -1.79 & 0.08 & 1.75 & 0.09 & RVS & Rg9 \\ \hline
        Rg9\_9 & 3828090710108683008 & 151.8146570289521 & -4.361546612412707 & 10.84 & 2679.0 & CFHT/ESPaDOnS & 136.2 & 4991.43 & 110.105 & 2.303 & 0.033 & -1.32 & 0.09 & 1.5 & 0.08 & RVS & Rg9 \\ \hline
        Rg9\_10 & 3830580489765691008 & 155.20250049439943 & -0.8941296684578997 & 12.35 & 4380.0 & VLT/UVES & 83.6 & 5047.768 & 104.281 & 2.355 & 0.035 & -1.91 & 0.07 & 1.62 & 0.12 & RVS & Rg9 \\ \hline
        Rg8\_5 & 5398944422751362560 & 170.7916124108969 & -35.39725640334401 & 11.87 & 2400.0 & VLT/UVES & 73.3 & 4589.391 & 82.035 & 1.373 & 0.044 & -1.86 & 0.08 & 2.09 & 0.11 & RVS & Rg8 \\ \hline
        Rg8\_6 & 3570774890078547840 & 181.25163895699416 & -13.61451625309093 & 11.54 & 2160.0 & VLT/UVES & 77.8 & 5270.126 & 125.437 & 2.948 & 0.029 & -2.1 & 0.08 & 1.35 & 0.17 & RVS & Rg8 \\ \hline
        Rg9\_11 & 1518697038547076352 & 189.1957349175882 & 36.34882939905674 & 11.59 & 2679.0 & CFHT/ESPaDOnS & 122.8 & 4523.132 & 69.342 & 1.509 & 0.034 & -1.44 & 0.08 & 1.77 & 0.12 & RVS & Rg9 \\ \hline
        Rg8\_7 & 3622623945729949440 & 197.0785845118268 & -11.562913255033726 & 11.80 & 600.0 & VLT/UVES & 69.7 & 4562.901 & 79.465 & 1.467 & 0.039 & -1.64 & 0.1 & 1.73 & 0.08 & RVS & Rg8 \\ \hline
        Rg9\_12 & 3743225081213792768 & 201.8322907865556 & 14.333826203341292 & 11.51 & 2679.0 & CFHT/ESPaDOnS & 138.9 & 5097.172 & 112.671 & 2.674 & 0.029 & -1.51 & 0.1 & 1.35 & 0.17 & RVS & Rg9 \\ \hline
        Rg8\_8 & 3726623482130353920 & 205.6309484556671 & 11.141995400363651 & 12.17 & 4320.0 & VLT/UVES & 80.2 & 4931.942 & 89.542 & 2.296 & 0.031 & -1.33 & 0.1 & 1.38 & 0.2 & RVS & Rg8 \\ \hline
        Rg8\_9 & 6110665270141152384 & 208.66316528741723 & -41.462899701094045 & 10.95 & 720.0 & VLT/UVES & 63.2 & 5029.009 & 102.807 & 2.156 & 0.026 & -2.22 & 0.1 & 1.7 & 0.13 & RVS & Rg8 \\ \hline
        Rg8\_10 & 6288008528536935168 & 208.8097215996244 & -21.48035972305724 & 12.07 & 1200.0 & VLT/UVES & 78.9 & 5043.968 & 116.406 & 2.278 & 0.035 & -2.06 & 0.09 & 1.65 & 0.1 & RVS & Rg8 \\ \hline
        Rg8\_11 & 3720752850247991040 & 209.7524844461709 & 7.000884543710807 & 10.45 & 1440.0 & VLT/UVES & 69.6 & 4506.014 & 83.962 & 1.479 & 0.029 & -1.5 & 0.12 & 1.88 & 0.19 & RVS & Rg8 \\ \hline
        Rg9\_13 & 1231766834895266048 & 213.6343134622754 & 14.963318033056218 & 10.75 & 720.0 & VLT/UVES & 89.3 & 4922.893 & 96.202 & 1.741 & 0.031 & -2.39 & 0.07 & 2.1 & 0.09 & LAMOST VMP & Rg9 \\ \hline
        Rg9\_14 & 6118201739454237440 & 216.8977998802665 & -37.89842644645348 & 11.38 & 2160.0 & VLT/UVES & 95.4 & 5047.075 & 97.593 & 2.259 & 0.03 & -2.24 & 0.09 & 1.72 & 0.11 & RVS & Rg9 \\ \hline
        Rg9\_15 & 1172712679547133440 & 219.2042972350077 & 7.927032557807032 & 10.39 & 2160.0 & VLT/UVES & 68.8 & 4232.786 & 64.03 & 0.731 & 0.04 & -1.32 & 0.13 & 2.0 & 0.15 & RVS & Rg9 \\ \hline
        Rg9\_16 & 3651423694314288384 & 220.6532377075662 & -0.246539699075751 & 12.64 & 3800.0 & VLT/UVES & 86.9 & 5308.49 & 115.688 & 2.935 & 0.027 & -2.33 & 0.12 & 1.45 & 0.13 & LAMOST VMP & Rg9 \\ \hline
        Rg8\_12 & 1208500997053446528 & 228.1887777920977 & 17.263395799401795 & 13.37 & 4750.0 & VLT/UVES & 100.2 & 5258.052 & 129.44 & 2.671 & 0.037 & -2.33 & 0.01 & 1.62 & 0.14 & LAMOST VMP & Rg8 \\ \hline
        Rg9\_17 & 1215930294043293056 & 230.3878750526661 & 24.08890142941875 & 13.22 & 3800.0 & VLT/UVES & 94.9 & 5017.87 & 101.48 & 2.035 & 0.044 & -2.5 & 0.13 & 1.85 & 0.2 & LAMOST VMP & Rg9 \\ \hline
        Rg8\_13 & 4543104496041198464 & 259.7586881651319 & 13.974075058759668 & 12.59 & 5700.0 & VLT/UVES & 77.4 & 4975.092 & 97.57 & 2.267 & 0.027 & -2.46 & 0.11 & 1.7 & 0.08 & LAMOST VMP & Rg8 \\ \hline
        Rg8\_14 & 4488707193348903296 & 267.68475701452303 & 9.387478749560852 & 12.10 & 4320.0 & VLT/UVES & 89.4 & 4606.943 & 85.81 & 1.471 & 0.04 & -1.38 & 0.09 & 2.14 & 0.15 & RVS & Rg8 \\ \hline
        Rg9\_18 & 4074532055211991296 & 285.66162159548963 & -24.131224327833745 & 10.72 & 4320.0 & VLT/UVES & 115.1 & 5147.418 & 112.269 & 2.427 & 0.027 & -2.04 & 0.11 & 1.57 & 0.09 & RVS & Rg9 \\ \hline
        Rg8\_15 & 4241432938148259456 & 297.4504042295441 & 2.163825234971092 & 10.74 & 2160.0 & VLT/UVES & 82.6 & 4755.95 & 101.411 & 1.294 & 0.036 & -1.66 & 0.1 & 1.82 & 0.14 & RVS & Rg8 \\ \hline
        Rg8\_16 & 6901006361471541248 & 307.06659936642274 & -12.013854556049768 & 11.47 & 1440.0 & VLT/UVES & 90.2 & 4809.58 & 94.236 & 1.666 & 0.037 & -2.02 & 0.11 & 2.1 & 0.07 & RVS & Rg8 \\ \hline
        Rg9\_19 & 2701974646152054272 & 325.0497265798293 & 8.489441566515794 & 11.92 & 900.0 & Subaru/HDS & 138.6 & 5270.505 & 118.437 & 2.848 & 0.033 & -2.14 & 0.11 & 1.42 & 0.1 & RVS & Rg9 \\ \hline
\enddata
\end{deluxetable*}

\section{Analysis of Sample Stars}
\label{para}
To investigate the chemical origins of the Rg8 and Rg9 substructures, we performed a detailed abundance analysis for our sample stars. We first outline the methods of stellar atmospheric parameters derivation and equivalent widths (EWs) measurement. The abundance analysis procedure as well as the treatment of uncertainties are then described briefly. Finally, to place our results in a broader Galactic context, we introduce the literature samples representing the stellar halo and accreted dwarf galaxies that we will use for comparison.

\subsection{Program Sample: Rg8 and Rg9 stars}


\subsubsection{Stellar Parameters}
\label{sec:par}

We follow the recipes from \citet{Sestito+2023MNRAS.518.4557S} to derive stellar parameters by adopting photometry from Gaia EDR3 \citep{Gaia+2021A&A...649A...1G}, extinction from \citet{Green+2019ApJ...887...93G}, and distances derived from a Bayesian approach with Galactic disc and halo priors \citep[following][]{Bailer-Jones+2015PASP..127..994B,Sestito+2019MNRAS.484.2166S, Sestito+2023MNRAS.518.4557S}.
We derive effective temperatures (\tefft) from a colour-temperature relation similar to the Infrared Flux Method \citep{gon+2009A&A...497..497G}, adapted to the Gaia DR3 photometry \citep{Mucciarelli+2021A&A...653A..90M}.
We then use the Stefan--Boltzmann equation to compute the surface gravity \citep[\logg; e.g.,][]{Kraft+2003PASP..115..143K, Venn+2017MNRAS.466.3741V} and, following the procedure described in \citet{Sestito+2023MNRAS.518.4557S}, we use an iterative process between \tefft\ and \logg\ to achieve self-consistent atmospheric parameters.
Uncertainties on \tefft\ and \logg\ are estimated via Monte Carlo simulations, propagating the uncertainties on the photometry \citep{Gaia+2021A&A...649A...1G}, the distance, and the extinction \citep{Green+2019ApJ...887...93G}.

\subsubsection{Equivalent Width Measurements}
\begin{figure*}
\plotone{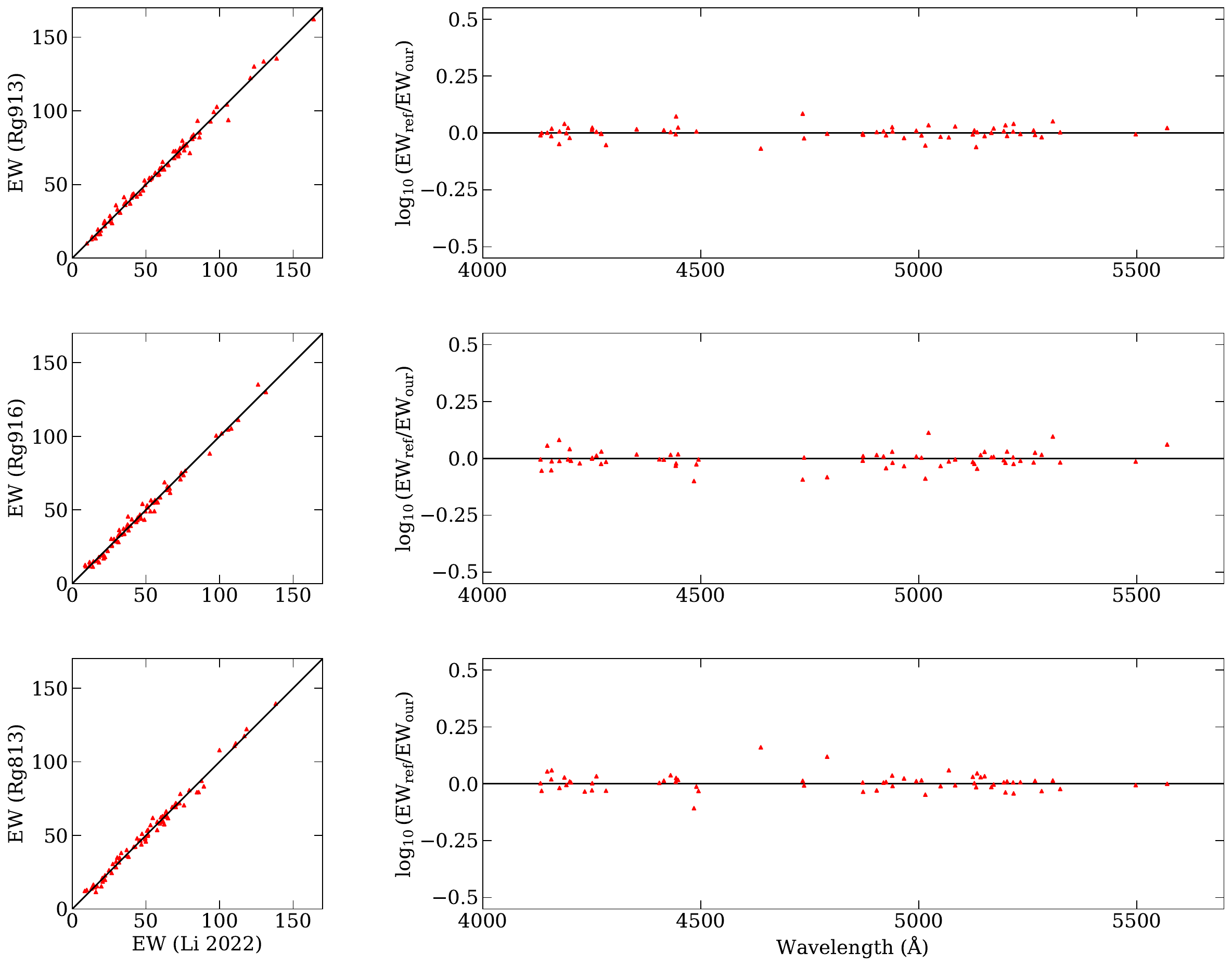}
\caption{
Comparison of our EW measurements with those of \citet{Li+2022} for three stars in common (each row corresponds to one star, as labelled). The average logarithmic difference of equivalent widths and the standard deviation are presented in the right panels. The results show no significant offsets, confirming the consistency of our EW measurements.
}

\label{con}
\end{figure*}

With the refined \logg\ and photometric \tefft, we derive metallicities ([Fe/H]) using the EWs of \ion{Fe}{2} absorption features. We choose \ion{Fe}{2} rather than \ion{Fe}{1} in this approach as the abundances for \ion{Fe}{2} are less affected by non-local thermodynamic equilibrium (NLTE) effects and temperature uncertainties
\citep[e.g.,][]{Amarsi+2016MNRAS.463.1518A}. The EWs were obtained by fitting Gaussian profiles using the IDL code TAME \citep{kang+2012}. We then calculate [Fe/H] using the MOOG code \citep{Sneden+1973} with 1D MARCS model atmospheres \citep{Gustafsson+2008}, assuming local thermodynamic equilibrium (LTE).
Finally, the microturbulent velocity ($\xi$) is determined by minimizing the trend between the \ion{Fe}{2} abundances and their reduced EW ($\log(\mathrm{EW}/\lambda)$). 

The stability and consistency of our EW measurements were checked using one star in our program that was observed more than once. The systematic offset between these multiple EW measurements is negligible. The scatter is typically less than 0.30~m\AA, which is comparable to the typical uncertainty of our EW measurements. This demonstrates the stability of our measurement procedure.

Furthermore, three other stars in our sample are in common with the study of \citet{Li+2022}. A comparison, shown in Figure~\ref{con}, demonstrates that our measured EWs are in agreement with their study. We find no significant offset in EWs, yielding a value of $|\langle\Delta\log\mathrm{EW}\rangle| / (\sigma(\Delta\log\mathrm{EW})/\sqrt{N}) < 5$. This value confirms the excellent agreement between our measurements and those from the literature.


\subsubsection{Abundance Analysis}
\label{sec:abun}
\begin{figure}
\plotone{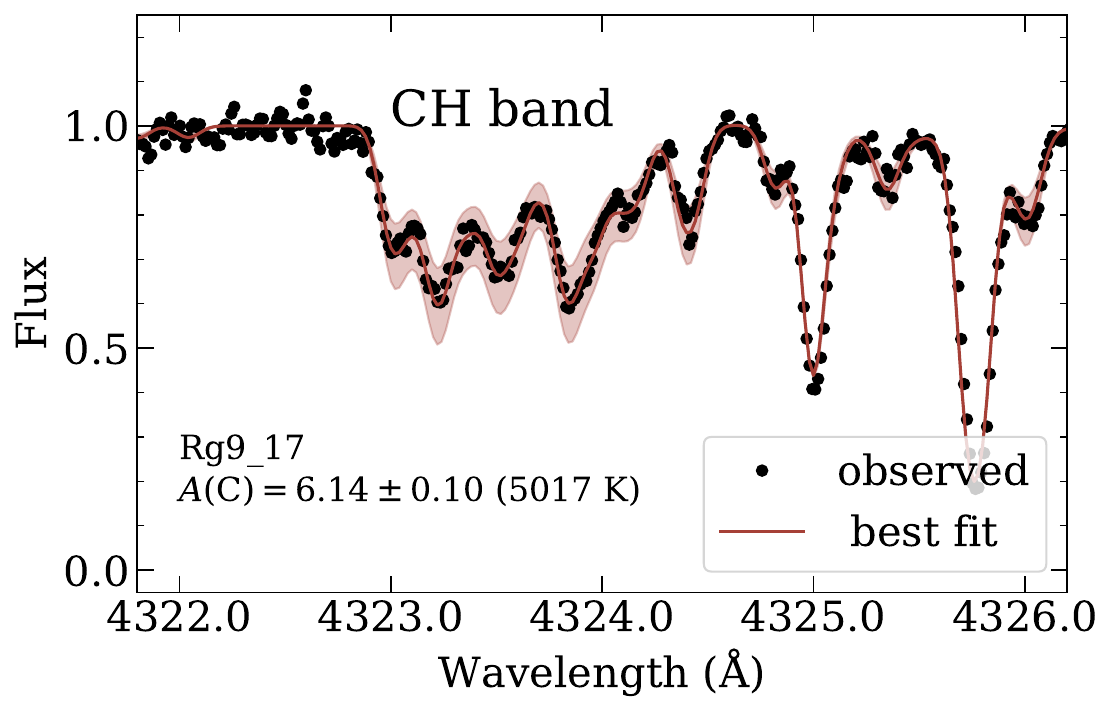}
\caption{Example of the spectral fitting of the CH 4323 \r{A} line for Rg9\_17 with LTE models. The observed spectrum is represented with black dots and the best-fitting result, with $A$(C) = 6.14, is shown as a solid line with a shaded uncertainty region.}
\label{CH}
\end{figure}

\begin{figure}
\plotone{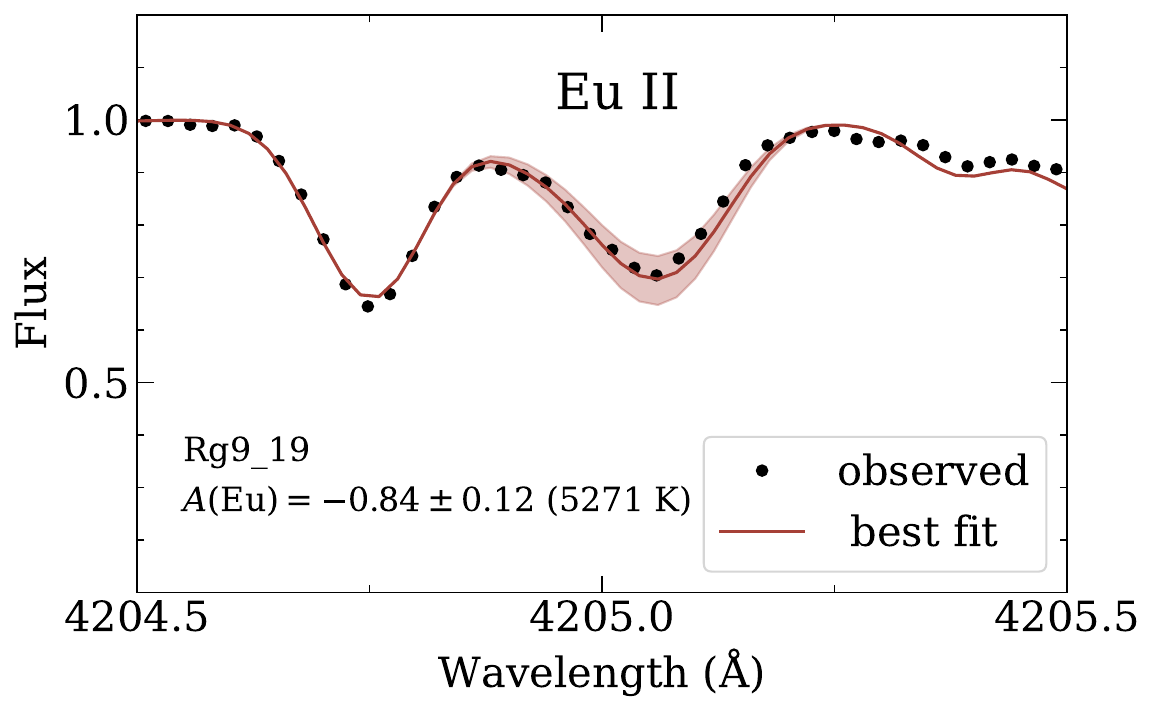}
\caption{Example of the spectral fitting of the Eu 4205 \r{A} line for Rg9\_19 with LTE models. Symbols are the same as in Figure~\ref{CH}, with the best-fitting result having $A$(Eu) = $-0.84\pm0.12$.}
\label{Eu}
\end{figure}

With EWs and stellar parameters determined, we now turn to the determination of elemental abundances for the program stars. For this analysis, we adopt the 1D plane-parallel, hydrostatic model atmospheres from the MARCS grid \citep{Gustafsson+2008}, assuming LTE. We use an updated version of the abundance analysis code MOOG \citep{Sneden+1973}, which treats continuous scattering as a source function that sums both absorption and scattering. The abundances are determined using two primary methods: analysis of EWs and spectral synthesis.

For 12 elements (Na, Mg, Si, Ca, Sc, Ti, Cr, Mn, Fe, Ni, Zn, and Y), abundances are computed from their measured EWs. The EWs are determined by fitting Gaussian profiles to single, unblended atomic absorption lines using the \texttt{IDL} code TAME \citep{kang+2012}. If an elemental abundance is derived from a single line, or if the derived abundances deviate by more than 3$\sigma$ from the average values computed for an atomic species from multiple lines, we confirm the abundances with spectral synthesis. For most of the lines used here, the synthetic spectra with abundances derived from the measured EWs match the observed spectra well. 

For other elements whose lines are blended or affected by hyper-fine structure (HFS), abundances are derived exclusively through spectral synthesis. These elements include C, Al, V, Co, Sr, Zr, Ba, La, and Eu. HFS and isotopic splitting are taken into account for Mn, Sr, Ba, La, and Eu. We show examples of the spectral synthesis fitting for the CH band and a Eu line in Figures \ref{CH} and \ref{Eu}, respectively.

Note that the above-mentioned analysis is expected to be subject to systematic uncertainties from NLTE effects. The photospheric solar abundances of \citet{Asplund+2009} were adopted when calculating the abundance ratios ([X/Fe]).
The abundance measurement results for our program sample are available in Zenodo and can be accessed via \textit{https://doi.org/10.5281/zenodo.17636850}.

\subsubsection{Systematic Uncertainties}
\label{sec:uncer}

The uncertainties on the derived abundances are estimated from two aspects: the uncertainties in the measurements and in the determination of the stellar parameters.
We estimate the measurement uncertainty from the standard deviation of abundances derived from individual lines ($\sigma_{\mathrm X}$) and the number of lines ($N_{\mathrm X}$), as $\sigma_{\mathrm X} / \sqrt{N_{\mathrm X}}$. 
For cases with only a single measured line ($N_{\mathrm X}=1$), the line-to-line scatter cannot be defined and we instead adopt a minimum random uncertainty representative of the data quality. This is typically $\sim 0.1$\,dex (derived from the standard deviation of abundances for elements with $N>10$ lines in the same spectrum). This estimate of the random uncertainties also includes the uncertainties on the $gf$ values of spectral lines.

We determine the uncertainties stemming from the uncertainties on the stellar parameters by individually varying \tefft, \logg, [Fe/H], and $\xi$ by their 1$\sigma$ uncertainties and determining the corresponding changes in the abundance value. To simplify the evaluation of these systematic errors, we assumed that their sources are independent, following the methodology adopted in previous studies \citep{Cseh+2018,Roriz+2021}. 
The total uncertainty was derived by adding these individual random and systematic uncertainties in quadrature. The final results for the entire sample, including the derived abundances, and the total uncertainty ($\sigma_{\mathrm{total}}$) are hosted on Zenodo and can be accessed via \textit{https://doi.org/10.5281/zenodo.17636850}.



\subsection{Literature Sample} 
To investigate the origin of substructures Rg8 and Rg9, we compare their chemical properties with two main literature samples: a broad sample of Galactic halo stars and a well-defined stellar stream whose progenitor is a dwarf galaxy.

For a benchmark of the Galactic halo population, we use the extensive dataset from \citet[][hereafter R14]{Roederer+2014AJ}. The \citetalias{Roederer+2014AJ} sample provides detailed and homogeneous abundances for 313 metal-poor stars. Given its large size and broad selection criteria, this sample is expected to be a composite of stars from numerous accreted progenitor systems with diverse properties, together with some in-situ stars formed in the Milky Way. It therefore effectively represents the chemically mixed and varied nature of the stellar halo.
To provide a clean comparison against a single, chemically simple accretion event, we use the data for the Cetus stream from \citet{Sitnova+2024A&A...690A.331S}, the first paper in the HR-GO series. The Cetus stream is a prime example of debris from a disrupted low-mass dwarf galaxy ($M_{\ast} \approx 10^6 M_{\odot}$). Its chemically homogeneous nature provides a clean template for an individual dwarf galaxy with mean metallicity [Fe/H] $\approx -2$.



\section{Elemental Abundances}
\label{sec:abund}
In this section, we present the detailed elemental abundances for 35 stars in Rg8 and Rg9 respectively, as they are originally identified.

\label{sec:res}
\begin{figure*}
\plotone{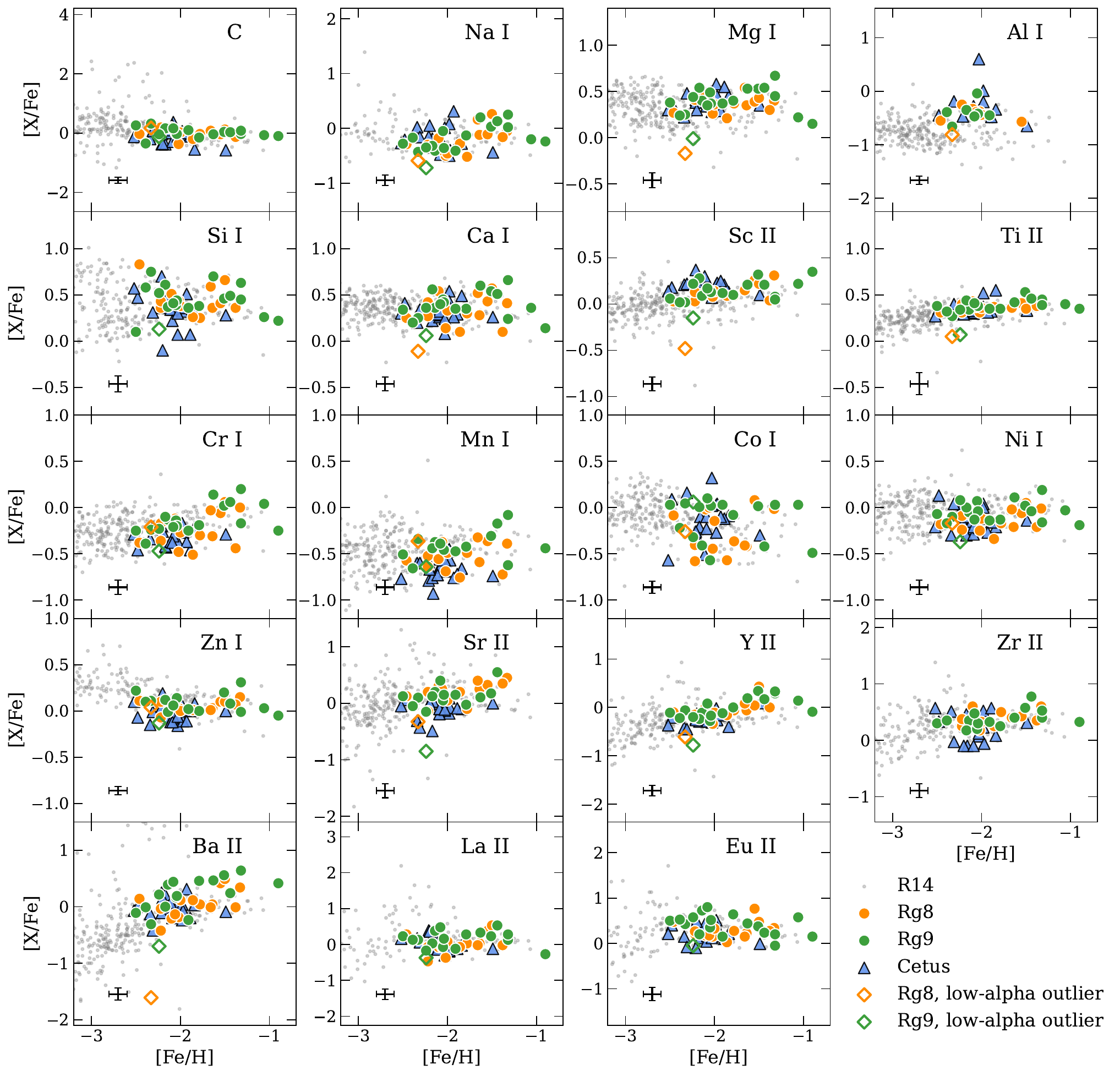}
\caption{
Chemical abundance ratios [X/Fe] as a function of metallicity [Fe/H] for our sample and comparison stars.
Our sample stars are shown as orange (Rg8) and green (Rg9) circles, with outliers indicated by empty diamonds of the same colors.
Literature comparison samples include blue triangles (Cetus) and gray dots (R14).
}
\label{fig:abundance}
\end{figure*}

\subsection{Carbon}
We determine carbon abundances from the CH band at around 4315 and 4322 \AA\ and we measure [C/Fe] ratios for 15 stars in Rg8 and 18 stars in Rg9. As our sample primarily consists of giant stars, the measured abundances require correction for carbon depletion effects from stellar evolution. We apply these corrections following the methodology of \citet{Placco2014} who model the CN cycle on the upper red-giant branch. This allows us to better account for the impact of stellar evolution on the carbon abundances in our analysis. The corrected [C/Fe] ratios are presented in Figure \ref{fig:abundance}. As shown in this figure, the two groups Rg8 and Rg9 are chemically indistinguishable in their carbon abundances. 
The median value for the Rg8 sample is [C/Fe] = $0.00$, which is consistent with the median value for the Rg9 sample ([C/Fe] = $0.02$). The two groups fully overlap in the [C/Fe]–[Fe/H] plane, showing no discernible offset or difference in their distributions. 
It is worth noting that to ensure a consistent comparison, we have applied the same \citet{Placco2014} corrections to the Cetus literature sample. Furthermore, the [C/Fe] ratios of our sample are fully consistent with those of the background halo stars (R14) and the Cetus substructure. The agreement suggests that Rg8 and Rg9 experienced a carbon enrichment history similar to that of the comparison halo samples.
We therefore conclude that Rg8 and Rg9 share an identical carbon enrichment history.

A notable characteristic of our sample is the complete absence of carbon-enhanced metal-poor (CEMP) stars, typically defined as [C/Fe]$>0.7$ \citep{Aoki+2007ApJ...655..492A}. This is significant, as CEMP stars are a common feature among other Galactic halo populations at these metallicities.

\subsection{$\alpha$-elements}
The $\alpha$-elements Mg, Si, Ca, and Ti provide further powerful constraints on the chemical enrichment history. 
The abundances for these elements are derived from multiple absorption lines of their neutral and/or singly-ionized species (e.g., \ion{Mg}{1}, \ion{Si}{1}, \ion{Ca}{1}, \ion{Ti}{1}, and \ion{Ti}{2}). 
We adopt the singly-ionized species (\ion{Ti}{2}) for Ti abundances. \ion{Ti}{2} is the dominant ionization species in metal-poor stars and is significantly less affected by non-LTE effects compared to \ion{Ti}{1}, offering a more reliable abundance indicator for this analysis
\citep[see e.g.,][]{Sitnova+2016MNRAS.461.1000S}.
These elements are primarily produced by core-collapse supernovae (CCSNe) and are thus excellent tracers of the early chemical enrichment history.

The [$\alpha$/Fe] ratios for Rg8 and Rg9 are presented in Figure~\ref{fig:abundance}. 
Both groups exhibit a consistent enhancement in all four $\alpha$-elements. A direct comparison of their median abundance ratios reveals a remarkable similarity. 
For Mg, we find median ratios of [Mg/Fe]$=0.39/0.40$ for Rg8/Rg9. 
Similarly, for Si the values are 0.41/0.45; for Ca, they are 0.33/0.36; and for Ti, they are 0.37/0.38.
The [$\alpha$/Fe] ratios of our sample are quite similar with those of the comparison samples. As shown in Figure~\ref{fig:abundance}. The R14 sample has relatively high [$\alpha$/Fe] very likely because it is substantially contributed by thick disk stars. The agreement with the Cetus sample across all four $\alpha$-elements without the appearance of the $\alpha$-knee suggests that these systems likely formed in environments with similar chemical enrichment histories.

The $\alpha$-element abundance of the two groups are in excellent agreement, showing no systematic offset or discernible difference in their scatter. We note that two stars at $[\mathrm{Fe/H}] \sim -1.0$ exhibit slightly lower ratios; however, given the typical measurement uncertainty ($\sim 0.1$\,dex), these deviations are insufficient to indicate a distinct feature. Furthermore, we find no robust evidence of an $\alpha$-knee in either group.
Based on a typical measurement uncertainty of $\sim$0.1\,dex for these elements, the mean values for Rg8 and Rg9 are statistically identical. 
This provides a powerful piece of evidence that the two substructures share a common origin and an identical history of $\alpha$-enrichment.

\subsection{Light Odd-Z Elements}
We also analyze the light odd-Z elements Na, Al, and Sc. For Na, we derive abundances using the Na doublet resonance lines at 5682 \AA\ and 5688 \AA. Al abundances are derived using the single \ion{Al}{1} resonance line at 3961 \AA, while Sc abundances are measured from two to eight \ion{Sc}{2} lines. To ensure consistent NLTE calculations, we adopt the most recent framework from \citet{Lind+2022AA...665A..33L}. These corrections for Na and Al are carried out after the spectroscopic analysis by interpolating the grids of observational atmospheric parameters and EWs, following \citep{Jiang+2025arXiv251005712J}.

Figure~\ref{fig:abundance} presents the abundance ratios for these elements. The median [Al/Fe] ratios, based on 7 stars in both groups, are nearly identical at $-0.39$ and $-0.42$ for Rg8 and Rg9, respectively. Similarly, the median [Sc/Fe] ratios are identical at 0.14 based on 15 (Rg8) and 18 (Rg9) stars. Both Al and Sc therefore show a remarkable consistency between the two groups. We note a quite larger offset of 0.13\,dex in the median [Na/Fe] values, which are 0.03 for Rg8 (15 stars) and 0.16 for Rg9 (14 stars). 
Although we exclude Na\,D resonance lines when measuring Na abundances, NLTE effects may also affect our Na measurements slightly. Since the Rg9 members are $\sim$ 0.1 dex more metal-poor than Rg8 members and the NLTE effects would be more significant for stars with lower metallicity, these differences may be due to the NLTE effects.
Previous studies have also revealed that a larger star-to-star scatter of approximately 0.2 dex is observed for Na abundances in very metal-poor stars compared to other elements \citep[e.g.,][]{Li+2022}. Therefore, although an offset exists in the Na values between Rg8 and Rg9, this difference is less significant when considering the large inherent scatter of the overall Na distribution.
Regarding the comparison with R14 sample and Cetus, we note that R14 exhibits wider spread in [Al/Fe], extending to lower values than observed in our stars. This difference is likely due to NLTE effects, given that the R14 values are based on LTE analysis. In contrast, Rg8 and Rg9 show good agreement with the Cetus in the [Al/Fe] space. For Na and Sc, our sample falls well within the distribution of both R14 and Cetus. The close resemblance to Cetus, combined with the general consistency with the halo trends, suggests a similar chemical evolution for the odd-Z elements.
The near-perfect agreement in Al and Sc, combined with the consistency seen in C and the $\alpha$-elements, provides a compelling body of evidence that Rg8 and Rg9 share a common origin.

\subsection{Iron-peak elements}
Our analysis includes the iron-peak elements V, Cr, Mn, Co, Ni, and Zn. We derive their abundances from a number of spectral lines, including three lines for V, seven for Mn, 11 for Cr, 15 for Ni, and two lines each for Co and Zn.

As shown in Figure~\ref{fig:abundance}, Rg8 and Rg9 exhibit highly similar abundance patterns across the iron-peak elements, with no significant differences observed between the two groups. For example, the median [Cr/Fe] values, based on 15 stars in Rg8 and 18 stars in Rg9, are $-0.26$ and $-0.18$, respectively. 
For nickel, the median [Ni/Fe] values are $-0.12$ for Rg8 and $-0.05$ for Rg9, measured from the same number of stars. The other iron-peak elements (V, Mn, Co, and Zn) also show a high degree of consistency between the two groups, with their median abundances agreeing well within the measurement uncertainties.

The iron-peak abundance ratios of our sample are indistinguishable from those of the R14 sample and the Cetus. Both Rg8 and Rg9 follow the general trends observed in these comparison samples, showing no significant deviations.
The consistent chemical signatures among the iron-peak elements provide additional, strong support for the conclusion that Rg8 and Rg9 are relics of the same progenitor system.

\subsection{Heavy elements}
\label{sec:heavy}

Lastly, we analyze the neutron-capture elements. The abundances for Sr, Zr, Ba, La, and Eu were derived using spectral synthesis based on 3 Sr\,\Rmnum{2} lines, 11 Zr\,\Rmnum{2} lines, 4 Ba\,\Rmnum{2} lines, 6 La\,\Rmnum{2} lines, and 6 Eu\,\Rmnum{2} lines, respectively. For Y, which has insignificant HFS \citep{Hannaford+1982}, abundances are calculated from the EWs of 10 Y\,\Rmnum{2} lines.

As presented in Figure~\ref{fig:abundance}, the abundance patterns of Rg8 and Rg9 are consistent across all six measured neutron-capture elements. This agreement is particularly notable for the heavier s-process elements; for example, the median [Sr/Fe] ratios are 0.21/0.13 for Rg8/Rg9, while the [Ba/Fe] ratios are similar at 0.04/0.21. The remaining elements (Zr, La, and Eu) show median abundances that are consistent between the two groups within measurement uncertainties.

When comparing with R14 and Cetus, we find that the R14 sample show a wider spread relative to our sample. This likely reflects the complex composition of the sample, which includes stars from diverse origins. Meanwhile, Rg8 and Rg9 exhibit a distribution that overlaps well with the Cetus. This agreement with Cetus further supports the notion that our stars originated from a specific accretion event.
The high degree of consistency across this diverse suite of elements provides strong evidence that Rg8 and Rg9 are fragments of the same progenitor.

\section{Results and Discussions}
\label{sec:result}
Rg8 and Rg9 are originally identified as two sets of retrograde DTGs that have very similar dynamical properties: relatively low energy and very retrograde orbits that likely belong to the same substructure \citep{Yuan+2020}. Our detailed analysis, spanning elements from carbon to neutron-capture products, reveals that the two groups are chemically indistinguishable, as shown in Fig.~\ref{fig:abundance}.
To quantify this homogeneity, we performed a Kolmogorov-Smirnov (KS) test on their metallicity distributions. The test yielded a $p$-value of 0.16 (statistic $D=0.35$), indicating no statistically significant difference between the two groups. Consistent statistical similarities are also found for other key chemical families, particularly the $\alpha$-elements (e.g., Mg, Si, Ca). This chemical similarity further supports that these retrograde DTGs share the same origin. Hereafter, we will treat these two groups as a single group and refer to it as the low-$E$ retrograde group.

As we have shown in Fig.~\ref{fig:dynamics}, the low-$E$ retrograde group strongly overlaps with the prominent retrograde substructure Thamnos, identified in \emph{Gaia} DR3 \citep{Dodd+2023A&A...670L...2D}. Our retrograde DTGs are originally identified from the VMP sample from \citep{Li+2018ApJS..238...16L}, whereas Thamnos was discovered based on orbital information determined using $Gaia$ RVS orbital properties, but without metallicity information \citep{Koppelman+2019}. Nevertheless, judging by their similar dynamical properties, it appears very likely that these retrograde groups identified differently are intrinsically the same structure. They could be an example of a single accretion event that is identified as different dynamical groups, as has been shown to happen in numerical simulations 
\citep[e.g.,][]{JeanBaptiste+2017A&A...604A.106J}. 
Moreover, several studies show that clusters identified in dynamical space can be created by in-situ stars, as the Milky Way disk is perturbed by passed merger events, as shown by  \citet{Orkney+2022MNRAS.517L.138O} and \citet{Thomas+2025arXiv250410398T} using the Auriga suite of cosmological magneto-hydrodynamic simulations
\citep{Grand+2017MNRAS.467..179G,Grand+2024MNRAS.532.1814G}. 

\subsection{Metallicity distribution function}
\label{mdf}
\begin{figure}
\plotone{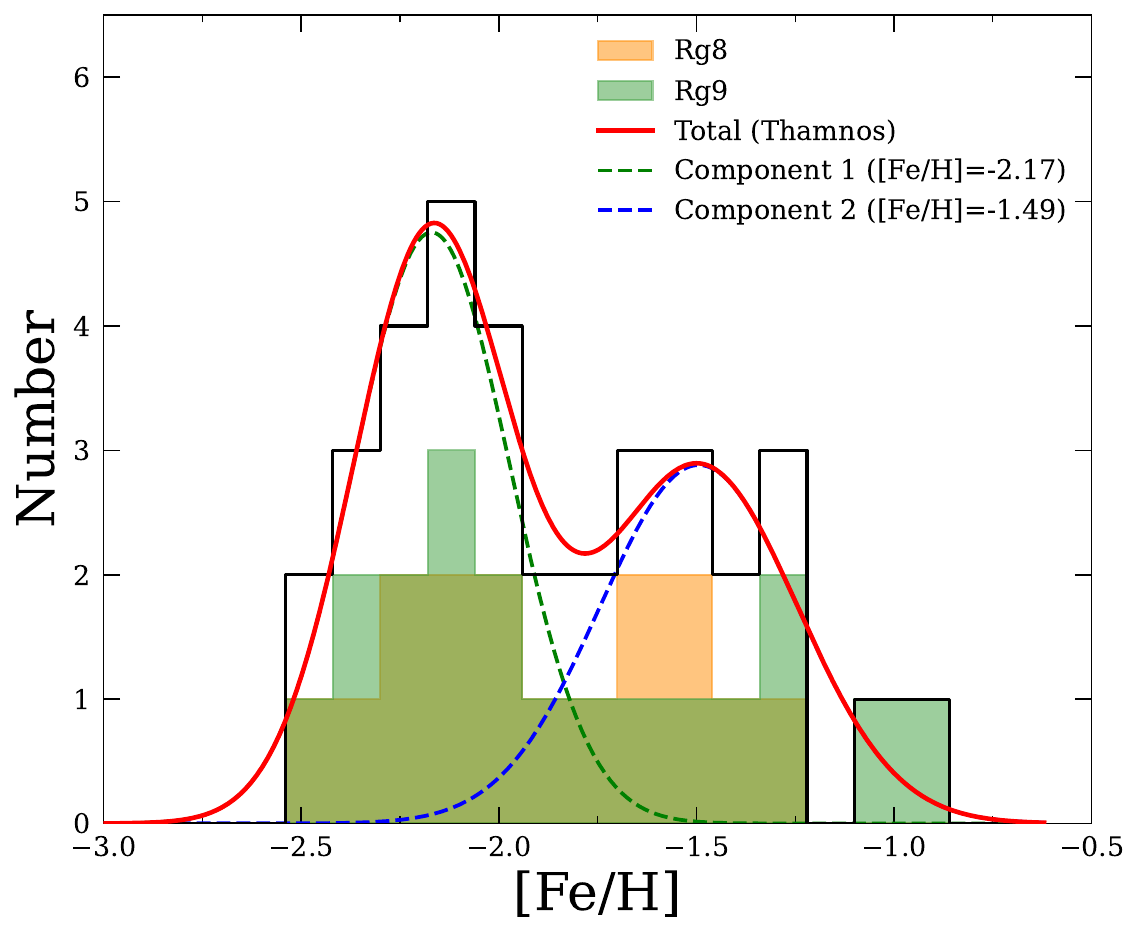} 
\caption{The MDF of the Thamnos sample. The individual histograms show the Rg8 (orange) and Rg9 (green) subgroups, which sum to the total distribution (black solid line). The distribution is fit by a double Gaussian model (red solid line). This fit reveals two components: Component 1 (green dashed line, with a peak at [Fe/H] = $-$2.17), representing the dominant progenitor, and Component 2 (blue dashed line, with a peak at [Fe/H] = $-$1.49).}
\label{fig:mdf}
\end{figure}

To further investigate if the low-$E$ retrograde group is from an accreted dwarf galaxy or in-situ, and if it has potential associations with Thamnos, we first study the metallicity distribution of Fe from our sample shown in Fig.~\ref{fig:mdf}. The metallicity distribution function (MDF) exhibits a clear bimodal distribution, characterized by a primary peak at [Fe/H]$ = -2.1$, and a secondary peak at [Fe/H] $= -1.5$. Using a double Gaussian decomposition, the more metal rich secondary component contributes to about 41\% of the whole sample. 

We emphasize that there was little knowledge about the metallicity of the 29 stars selected from \emph{Gaia} DR2 before our follow-up campaign. The fact that 13 of these stars have [Fe/H] $<-2.0$ is itself strong evidence that the low-$E$ retrograde group is dominated by a VMP population (19 out of 35 stars in total). One possible explanation of the metal rich component is the contamination from the in-situ MW stars; \eg the splashed disk stars perturbed by passed merger events \citep{Di+2019A&A...632A...4D,Gallart+2019NatAs...3..932G,Belokurov+2020MNRAS.494.3880B}. Simulation results show that such contamination can contribute to the retrograde halo at the 20--60\% level \citep{Thomas+2025arXiv250410398T}.

A similar bimodality of the MDF of the Thamnos substructure was recently reported by \citet[][hereafter \citetalias{Ceccarelli+2025arXiv251006332C}]{Ceccarelli+2025arXiv251006332C}, who followed this substructure as identified by \citet{Dodd+2023A&A...670L...2D}. \citetalias{Ceccarelli+2025arXiv251006332C} found a main metal rich component that peaks at [Fe/H] $= -1.5$, as well as a bump at [Fe/H] $= -2.1$. The peak metallicities of these two components coincide with our findings, but we note that the metal-rich population dominates the \citetalias{Ceccarelli+2025arXiv251006332C} sample (78\%) more significantly than that in our sample (41\%). \citetalias{Ceccarelli+2025arXiv251006332C} interpret the low-metallicity component as the potential signature of the Thamnos substructure. If we define Thamnos as the debris from an accreted dwarf galaxy, the VMP population that dominates the low-$E$ retrograde group in this work provides strong evidence for its existence.

We conclude that the low-$E$ retrograde group is composed of the Thamnos substructure, contaminated by the metal-rich in-situ stars. Using the prominent MDF peak and the mass-metallicity relation for dwarf galaxies \citep{Kirby+2013ApJ...779..102K}, we estimate a progenitor stellar mass of $\sim 10^6 \, M_{\odot}$. The Thamnos structure identified in this work exhibits lower contamination from in-situ stars, yet occupies a wider distribution in dynamical space compared to that from \citet{Dodd+2023A&A...670L...2D} as shown in Fig.~\ref{fig:dynamics}. We attribute this to the fact that our DTGs are originally identified from the VMP sample using a \texttt{STARGO} applied to a 4-dimensional dynamical space ($E$, $L$, $\theta$, $\phi$) via self-organizing maps \citep[SOM; see details in][]{Yuan+2018ApJ...863...26Y}, with the latter two angles characterizing the orbital direction \citep{Yuan+2020}. The SOM capture the latent structural relationships of these DTGs and ensures robust membership identification even when expanding the sample to the full \emph{Gaia} RVS catalog.

\subsection{The absence of an $\alpha$ knee}
\label{alpha}
\begin{figure}
\includegraphics[width=0.45\textwidth]{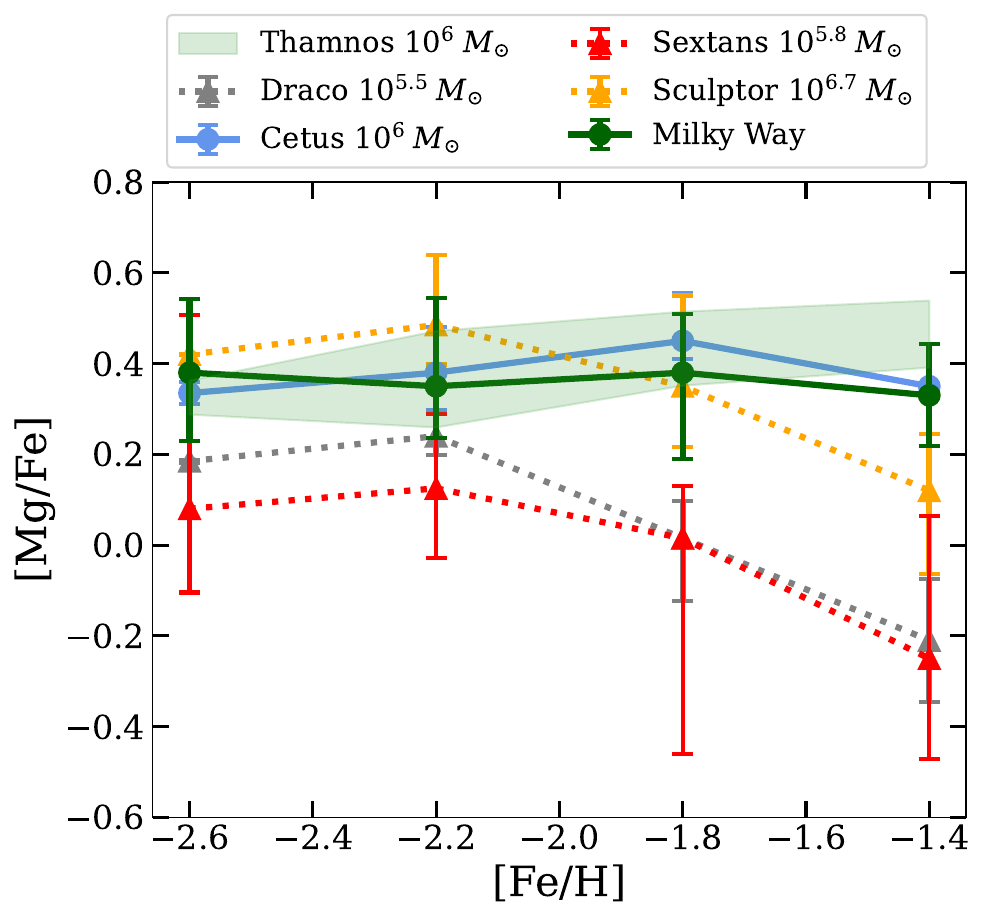}
\caption{[Mg/Fe] for known dwarf galaxies compared to Thamnos, which is represented by a green region.
In order from highest to lowest mass, abundances are plotted for Sculptor
\citep{Hill+2019A&A...626A..15H}
(dotted skyblue line),
Sextans
\citep{Theler+2020A&A...642A.176T}
(dotted red line),
Draco
\citep{Yamada+2013, suda+2017}
(dotted grey line), 
halo stars (R14) stars
\citep{Roederer+2014AJ}
(darkgreen line)
and Cetus stars
\citep{Sitnova+2024A&A...690A.331S}.
Stellar masses are given in the legend.}
\label{dwarf}
\end{figure}

We now interpret the abundance results within the framework that Thamnos represents a low-mass accreted dwarf galaxy. The [$\alpha$/Fe] ratio serves as a powerful chemical clock: the downturn point, or `$\alpha$-knee,' that is observed in many dwarf galaxies and in the Milky Way marks the epoch when Type Ia supernovae (SNe Ia) began to contribute significantly to chemical enrichment. Simulations of Milky Way-like galaxies demonstrate that [$\alpha$/Fe] is a crucial diagnostic for distinguishing in-situ from accreted populations as this quantity is highly sensitive to the star formation efficiency of the birth environment \citep{Khoperskov2023, Mason+2024MNRAS.533..184M,Thomas+2025arXiv250410398T}.

As shown in Figure~\ref{fig:abundance}, our sample defines an [$\alpha$/Fe] plateau at $\approx0.4$. For comparison, Fig.~\ref{dwarf} summarizes the trend of [Mg/Fe] vs. [Fe/H] for Thamnos, as well as for three dwarf galaxies with mean metallicity [Fe/H] $\approx-2.0$: Draco \citep{Cohen+2009ApJ...701.1053C}, Sextans
\citep{Theler+2020A&A...642A.176T}, and Sculptor \citep{Hill+2019A&A...626A..15H}. The absence of an $\alpha$ knee in Thamnos is clearly distinct from the surviving low-mass dwarf galaxies, whose ``$\alpha$ knee'' appears at [Fe/H]$\lesssim-2.0$. Notably, the two most metal-rich stars in our sample, with [Fe/H] $\sim -1.0$, show slightly lower [$\alpha$/Fe] ($\approx$\, 0.2) in Fig.~\ref{fig:abundance}. This is consistent with the ``$\alpha$ knee'' from the splashed disk stars instead of that expected in dwarf galaxies.

In Fig.~\ref{dwarf}, we also include the mean track of the Milky Way sample from \citet{Roederer+2014AJ}. In order to ensure a fair comparison, we account for potential offsets between turn-off stars and giants at low metallicities \citep{Li+2022, Roederer+2014AJ}. As our Thamnos sample and the dwarf galaxy samples are dominated by giants, we compare them specifically to the giant star subsample of \citetalias{Roederer+2014AJ}. The flat trend of the Milky Way sample is mainly due to the significant contribution from the in-situ stars that show an $\alpha$ knee at [Fe/H] $\approx-1.1$.

\begin{figure*}
\plotone{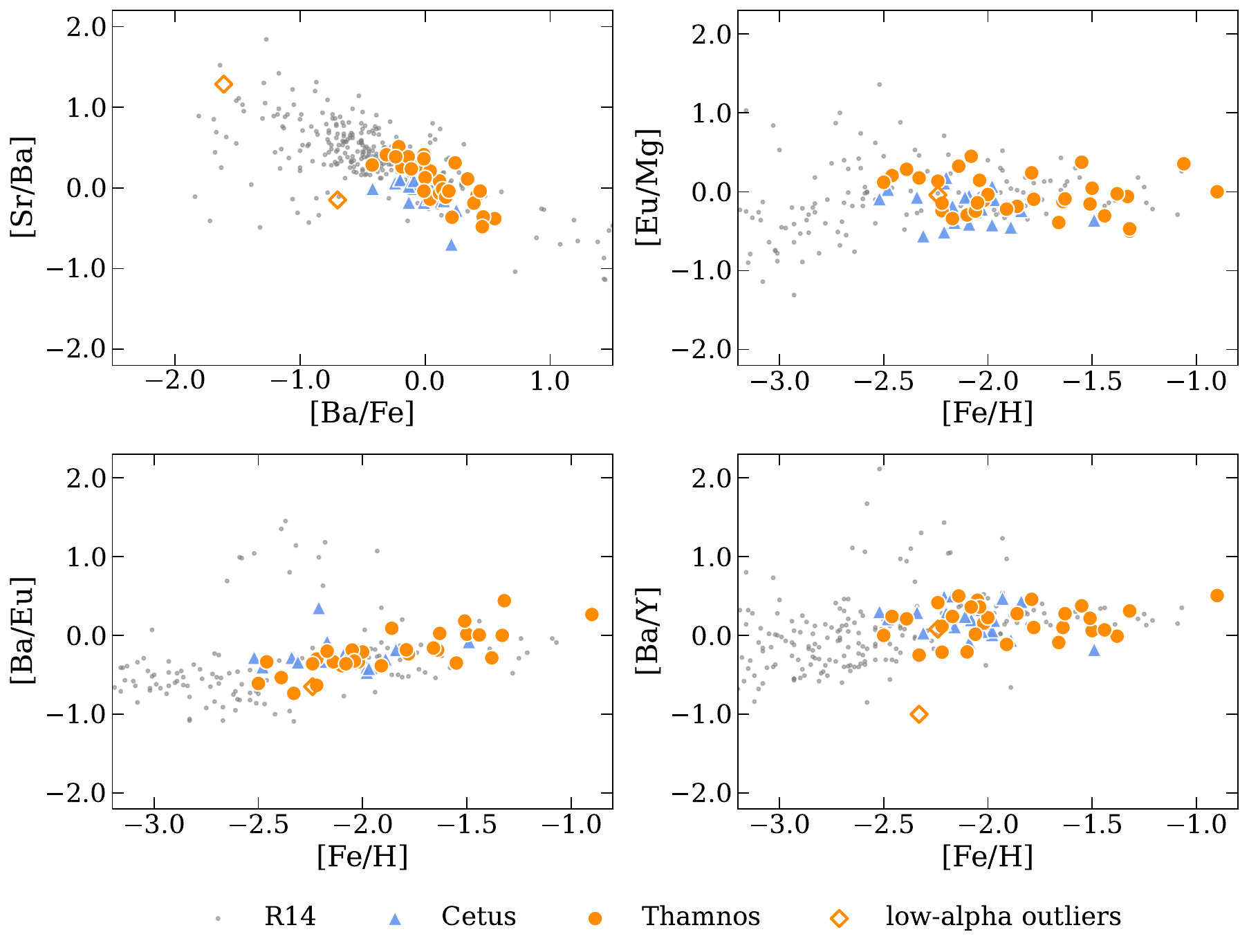}
\caption{Neutron-capture abundance ratios for our sample stars.
The panels show [Eu/Mg] vs. [Fe/H] (top left), [Ba/Eu] vs. [Fe/H] (top right), [Ba/Y] vs. [Fe/H] (bottom left), and [Sr/Ba] vs. [Ba/Fe] (bottom right).
Our Thamnos sample (orange circles) and Cetus members (blue triangles) are shown. 
The two outliers are marked as orange diamonds.
The background comparison sample from halo stars (R14) is shown as gray dots.}
\label{neu}
\end{figure*}

However, for halo substructure, a more fair comparison is obtained by comparing our sample to a disrupted dwarf galaxy instead of surviving ones that have likely seen their stellar evolution proceed for longer. We do so by comparing our sample with the  Cetus stream, which is undoubtedly the debris from an accreted dwarf galaxy. This system has a simple MDF that, like Thamnos, peaks at the same metallicity of [Fe/H] $= -2.1$. Strikingly, the Cetus stream also does not show an $\alpha$ knee \citep{Sitnova+2024A&A...690A.331S}, as shown in Fig.~\ref{dwarf}. This feature can be explained by a truncated star formation history. The Cetus progenitor very likely fell into the Milky Way potential more than 8 Gyr ago, as inferred from N-body simulations 
\citep{Chang+2020ApJ...905..100C}. Therefore, its star formation probably stopped before Type Ia SNe could play an important role. Thamnos has a lower orbital energy than the Cetus stream, which suggests that it fell in even earlier than Cetus. The similarly truncated star-formation history as well as the absence of an $\alpha$ knee are then natural outcome of this scenario.

The absence of an $\alpha$ knee for Thamnos has also been shown by \citet{Koppelman+2019, Monty+2020, Horta+2023,Zhang+2024}. The recent HR spectroscopic study from \citetalias{Ceccarelli+2025arXiv251006332C} reveals that the majority of the Thamnos stars have rather enhanced [$\alpha$/Fe] and that a hint of an $\alpha$ knee is only present at lower metallicity, around [Fe/H] $\approx -2.3$. We do not see such a feature in our sample. However, there are two VMP stars with [Fe/H] $= -2.33$ and $-2.24$ that exhibit unusually low [Mg/Fe] ratios, which make them two outliers in Figure~\ref{fig:abundance}. While most Thamnos stars have [Mg/Fe] $\sim$ 0.4--0.5, these two stars fall significantly below that trend, with [Mg/Fe] $= -0.17$ and $-0.01$. These values are even lower than the stars in C25 ([Mg/Fe] $> 0.1$) that are considered as the sign of a potential $\alpha$ knee. Besides Mg, elements like Si, Ca, Ti, and Sc also exhibit slightly lower abundance ratio with respect to Fe for these two stars. Clearly, these two outliers deviate from the main track of the other VMP Thamnos stars in the chemical space.

While the iron-peak abundances of the outliers are unremarkable, their neutron-capture element patterns show distinct features as well. The two outliers are clearly offset from the trend of this main Thamnos group that follows the track of halo stars shown in the upper-left panel of Figure~\ref{neu}. Instead, they resemble the chemical signatures widely observed in UFD galaxies 
\citep{Simon+2019ARA&A..57..375S,Ji+2019ApJ...870...83J,Frebel+2014ApJ...786...74F}. Compared to more massive dwarf galaxies as well as halo stars, these low-mass systems are less enriched in neutron-capture elements produced by core-collapse supernovae (CCSNe) \citep[see \eg][]{Ji+2016ApJ...830...93J}.
These two stars could also have been born from gas polluted by a CCSN that underwent negligible fallback of the innermost ejecta enriched in Fe-peak elements. Ejecta from such CCSNe can naturally result in sub-solar [Mg/Fe] along with lower values of [X/Fe] for Si, Ca, Sc, and Ti  \citep{Jeena+2025ApJ...981...55J}.

As we can see from Fig.~\ref{fig:dynamics}, these two low-$\alpha$ outliers remain consistent with the other Thamnos stars in the dynamical space. By combining their chemo-dynamical properties, we propose that they could be the relics of stellar systems with even lower masses, such as ultra-faint dwarf galaxies \citep[UFD; see e.g.,][]{Frebel+2018ARNPS..68..237F,Simon+2019ARA&A..57..375S,Ji+2019ApJ...870...83J,Ou+2025AJ....169..279O,Atzberger+2025OJAp....8E..68A}. Their progenitor systems have substantial SNe Ia contributions much earlier than when they reach [Fe/H] $\approx-2.0$. The exceptionally low [$\alpha$/Fe] could be the outcome of a top-light IMF that can make the enrichment from SNe\,Ia more significant than CCSNe, based on the chemical evolution model of Bo\"{o}tes I \citep{Yan+2020A&A...637A..68Y}. 
The progenitor UFD of these two outliers could have merged with the Thamnos progenitor or simply fell into the MW with very similar orbits and time as the Thamnos progenitor (for instance, if they were satellites of the Thamnos progenitor). 
Alternatively, they may have been accreted at a different epoch from a distinct progenitor, and simply overlap dynamically with the Rg8-9 structure by chance.


\subsection{More evidence from other elements}

In addition to the $\alpha$ elements, other abundance tracers also show a flat trend in Thamnos, as shown in Fig.~\ref{fig:abundance}. This is similar to what is seen in the Cetus stream \citep{Sitnova+2024A&A...690A.331S} and adds weight to the scenario of a short star-formation timescale. Sc is mainly produced in CCSNe \citep{Kobayashi+2020} and its abundance shows no sign of decline over the observed metallicity range. Similarly, both [Ni/Fe] and [Zn/Fe] remain nearly constant down to [Fe/H] $\approx-1.0$. Since SNe\,Ia produce little Zn \citep{Iwamoto+1999} and tend to lower [Ni/Fe] in accreted populations after their onset \citep{nissen+2011}, the absence of any decline in these elements indicates that Thamnos experienced a rapid, CCSN-dominated enrichment with little or no contribution from SNe\,Ia.

The trends of key neutron-capture elements (Sr, Br, Eu) provide detailed insight into the star formation and enrichment timeline of the dwarf galaxies shown in Figure~\ref{neu}. The [Eu/Mg] ratio (upper-right panel of Figure~\ref{neu}) exhibits a flat trend across the entire metallicity range studied here. 
This feature indicates that the rate of Eu (an r-process dominated element) and Mg (a pure $\alpha$ element) is similar to that produced in massive stars. Crucially, the absence of a rising [Eu/Mg] trend reveals no significant contribution from time-delayed r-process sources, such as neutron star mergers. This behavior at [Eu/Mg]$\sim0.0$\,dex is similar to that of the high-$\alpha$ population in the Galactic halo and Cetus.

In contrast, the [Ba/Eu] ratio clearly shows an increasing trend with metallicity (bottom-left panel of Figure~\ref{neu}). Since Eu is a pure r-process element, while Ba is produced by both r  and s processes, this trend reflects that the star-formation timescale allows for sufficient s-process enrichment from AGB contributions. [Ba/Y], shown in the bottom right panel of Figure~\ref{neu}, traces the relative contribution of heavy s-process elements (like Ba) to light s-process elements (like Y). As demonstrated by \citet{Fenner+2006ApJ...646..184F}, systems with suppressed or truncated star-formation histories are expected to show flat [Ba/Y] trends at low metallicities, because the delayed enrichment from low-mass AGB stars is cut off. Our observation of a flat [Ba/Y] trend in Thamnos strongly suggests that its progenitor had a short star-formation timescale, consistent with our findings from the $\alpha$ elements. Furthermore, the specific abundance pattern of [Ba/Y] $\sim 0.0$, combined with substantial Ba enrichment is well reproduced by s-process nucleosynthesis in rapidly rotating massive stars \citep{Banerjee+2018ApJ...865..120B}. Crucially, since these contributors are massive, short-lived stars, this scenario remains fully consistent with the conclusion of a quick star-formation history truncated before the onset of significant low-mass AGB contribution. This scenario is also similar to what is observed in the Cetus stream (blue triangles in Figure~\ref{neu}) that displays an increasing [Ba/Eu] ratio accompanied by a flat [Ba/Y] sequence. The remarkable similarity between these two stellar systems indicates that both progenitor dwarf galaxies have experienced truncated star formation.

Notably, there are no CEMP giants detected in our sample that is selected purely based on dynamics. This lack of CEMP stars is similar to what is seen in the Cetus stream and agrees with the analysis of \citet{Zhang+2024}, who also reported a lack of CEMP giants associated with Thamnos. However, it is worth noting that \citet{Zhang+2024} detected a significant CEMP fraction ($\sim$20\%) among the turnoff members of Thamnos. Given that the fraction of CEMP stars is known to be significantly lower in giants compared to turnoff stars \citep[e.g.,][]{Li+2022}, the absence of CEMP stars in our sample is likely attributed to the evolutionary status of our targets (all giants) rather than an intrinsic deficiency of carbon in the Thamnos progenitor.

\section{Conclusions}
\label{sec:con}
In this work, we expand two sets of low-$E$ retrograde DTGs (Rg8 \& Rg9) from \citet{Yuan+2020}, originally identified from the LAMOST VMP sample \citep{Li+2018ApJS..238...16L}, to the entire \emph{Gaia} RVS sample without metallicity constraints.
Notably, these groups occupy a similar region in dynamical space to Thamnos.
We obtain high signal-to-noise, HR spectra of 35 stars from these two DTGs, including six stars selected from the VMP sample. Despite these stars being mainly selected from their dynamical properties, independently of their metallicities, we find that the majority of them (19) have [Fe/H]$<-2.0$. This gives strong evidence that these DTGs are dominated by VMP stars. We derive detailed elemental abundances for these 35 stars, and summarize our main findings below. 

\begin{enumerate}
    \item We show that Rg8 and Rg9 are chemically indistinguishable, which strongly suggests that they share a common origin, in-line with their similar dynamical properties. We therefore combine them into a unified low-$E$ retrograde group to investigate the detailed chemical properties of Thamnos.
 
    \item The MDF of the low-$E$ retrograde group has a dominant peak at [Fe/H] = $-2.1$, and a secondary peak at [Fe/H] = $-1.5$. The group largely overlaps with the Thamnos substructure identified in dynamical space by \citet{Dodd+2023A&A...670L...2D}, leading us to assume that they are the same structure. Thamnos suffers from significant contamination from more metal-rich, in-situ MW stars \citep{Ceccarelli+2025arXiv251006332C} but we identify the dominant low-metallicity population of our group as representative of the true Thamnos stellar population. The fact that it is so metal-poor provides strong evidence for the existence of Thamnos as an accreted substructure. From the mean metallicity of the VMP population, we estimate that the Thamnos progenitor had a stellar mass $\approx10^6M_{\odot}$.

    \item We find no sign of an $\alpha$ knee in the Thamnos abundances analyzed here, which is consistent with previous studies \citep{Koppelman+2019, Horta+2023}. This behavior, as well as the mean metallicity of Thamnos, is similar to what is observed in the Cetus stream that undoubtedly comes from an accreted dwarf galaxy. A possible scenario to explain these observations is that the progenitors of these streams had their star formation stopped before Ia supernovae played an important role. This is further supported by the flat trend seen for other $\alpha$ elements in both systems.

    \item The increasing trend of [Ba/Eu] shows that Thamnos has been continuously enriched by $s$-process elements. The relatively flat [Ba/Y] trend indicates that the contribution from low-mass AGB stars is negligible. This feature further supports that the Thamnos progenitor underwent only a short period of star formation.

\end{enumerate}

In conclusion, our analysis establishes Thamnos as the chemical relic of a low-mass dwarf galaxy that was accreted early into the Milky Way halo. Its chemical signature distinguishes it from classical dwarf galaxies like Sculptor. It resembles that of the dwarf-galaxy stream Cetus and points to a relatively short star-formation episode that was truncated upon the accretion of the Thamnos progenitor.

\section{Data Availability}
The dynamical information and abundance measurement results for our program sample are available in Zenodo and can be accessed via \textit{https://doi.org/10.5281/zenodo.17636850}.

\begin{acknowledgments}
This work is supported by the Strategic Priority Research Program of Chinese Academy of Sciences grant Nos. XDB1160103, the National Key $R\&D$ Program of China Nos. 2024YFA1611903, 2023YFE0107800 and 2024YFA1611601, the National Natural Science Foundation of China grant Nos. 12588202 and 12222305, the science research grants from the China Manned Space Project and the China Manned Space Program with grant No. CMS-CSST-2025-A12. This work is supported by the CAS Project for Young Scientists in Basic Research (No. YSBR-092).
This work was supported by the ``action thématique'' Cosmology-Galaxies (ATCG) of the CNRS/INSU PN Astro. 
TM is supported by a Gliese Fellowship at the Zentrum f\"ur Astronomie, University of Heidelberg, Germany.

We thank the ESO staff for carrying out the VLT-UT2 observations and reducing the data. Allocation of VLT Director’s Discretionary Time is gratefully acknowledged. Based on observations collected at the European Organisation for Astronomical Research in the Southern Hemisphere under ESO programme(s) 106.21M6.001.
This work is based on observations obtained at the Canada-France-Hawaii Telescope (CFHT) which is operated by the National Research Council of Canada, the Institut National des Sciences de l'Univers of the Centre National de la Recherche Scientifique of France, and the University of Hawaii.
This research is based in part on data collected at the Subaru Telescope, which is operated by the National Astronomical Observatory of Japan. This paper also includes data gathered with the 6.5 m Magellan Telescopes, located at Las Campanas Observatory, Chile. 
This work has made use of data from the European Space Agency (ESA) mission {\it Gaia} (\url{https://www.cosmos.esa.int/gaia}), processed by the {\it Gaia} Data Processing and Analysis Consortium (DPAC,
\url{https://www.cosmos.esa.int/web/gaia/dpac/consortium}). Funding for the DPAC has been provided by national institutions, in particular the institutions participating in the {\it Gaia} Multilateral Agreement.

\end{acknowledgments}

\begin{contribution}



\end{contribution}

%
\facilities{Subaru/HDS, VLT/UVES, CFHT/ESPaDOnS, Magellan/MIKE}

\software{
TOPCAT \citep{Taylor+2005ASPC..347...29T}  }


\appendix

\bibliography{Rg89}{}
\bibliographystyle{aasjournalv7}



\end{document}